%% file: main.tex
\documentclass{article}
\usepackage{spconf,amsmath,graphicx}
\usepackage{tikz,pgfplots}

\input{preamble}
\usepackage{xurl}

\newcommand{\na}{\mathsf{n \backslash a}}
\title{
Coding For The Unsourced A-channel with Erasures: \\
the Linked Loop Code
}
\name{
William W. Zheng, 
Jamison R. Ebert, 
Stefano Rini, 
and Jean-Francois Chamberland
\thanks{
This material is based upon work supported, in part, by the National Science Foundation (NSF) under Grants CCF-2131106 and CNS-2148354, by Qualcomm Technologies, Inc., through their University Relations Program, and by the Ministry of Science and Technology (MOST) under Grant 110-2221-E-A49-052.
William Weijia Zheng is with the Department of Information Engineering, the Chinese University of Hong Kong, HKSAR (email: wjzheng@link.cuhk.edu.hk).
J. R. Ebert and J.-F. Chamberland are with the Department of Electrical and Computer Engineering, Texas A\&M University, College Station, TX 77843, USA (emails: \{jrebert,chmbrlnd\}@tamu.edu).
S. Rini is with the Department of Electrical and Computer Engineering, National Yang Ming Chiao Tung University (NYCU), Hsinchu, 300, TW (email: stefano.rini@nycu.edu.tw).}
}
\address{}

\begin{document}

\maketitle

\begin{abstract}
%
The A-channel is a noiseless multiple access channel in which users simultaneously transmit $Q$-ary symbols and the receiver observes the set of transmitted symbols, but not their multiplicities. 
An A-channel is said to be unsourced if, additionally, users’ transmissions are encoded across time using a common codebook and decoding of the transmitted messages is done without regard to the identities of the active users \cite{lancho2022finite}.
An interesting variant of the unsourced A-channel is the unsourced A-channel with erasures (UACE), in which transmitted symbols are erased with a given independent and identically distributed probability. 
In this paper, we focus on designing a code that enables a list of transmitted codewords to be recovered despite the erasures of some of the transmitted symbols. 
To this end, we propose the linked-loop code (LLC), which uses parity bits to link each symbol to the previous $M$ symbols in a tail-biting manner, i.e., the first symbols of the transmission are linked to the last ones. 
The decoding process occurs in two phases: the first phase decodes the codewords that do not suffer from any erasures, and the second phase attempts to recover the erased symbols using the available parities. 
We compare the performance of the LLC over the UACE with other codes in the literature and argue for the effectiveness of the construction. 
Our motivation for studying the UACE comes from its relevance in machine-type communication and coded compressed sensing.

\end{abstract}

\section{Introduction}
    The A-channel with erasures is defined by the input/output relationship 
    \ea{
    \textstyle Y = \bigcup_{k \in [K]}  \delta_{E_{k}} (x_{k}),
    \label{eq:channel error}
    }
    where the $k$th users's symbol $x_{k} \in \Qcal, |\Qcal|=Q$,
    %
    the erasure indicator $E_{k} \stackrel{\text{iid}}{\sim} \Bcal(p_e)$, and the erasure function is defined as 
    \ea{
    \delta_{E}(x) \coloneqq \lcb \p{
    x  &  E = 0  \\
    \emptyset & E = 1.
    }\rnone
    }
    Above, $\Bcal(p)$ indicates the Bernoulli distribution with parameter $p_e \in [0,1]$.
    %
    %
In the A-channel with erasures of \eqref{eq:channel error},
each symbol is received by the receiver with probability $1-p_e$ and erased with probability $p_e$. 
%
    %
%
The unsourced A-channel with erasures (UACE) is similar to the model described in \eqref{eq:channel error}; however, the receiver is only interested in identifying the transmitted codewords, rather than the traditional source-message pairs. 
%
In this article, we introduce the UACE paradigm, and we propose a coding scheme for the challenge which we term linked-loop code (LLC). 
The LLC uses parity bits to link coded symbols in a tail-biting manner, thus allowing for codewords to be recovered even when some of their symbols have been erased.
We evaluate the performance of the LLC and compare it to that of other coding schemes available in the literature.

\noindent
\textbf{Literature Review:}
The A-channel is described in \cite{chang1981t}, where it models a multiple-access channel that employs frequency division for transmission without the availability of signal intensity. 
The authors of \cite{chang1981t} determine the mutual information of the A-channel under uniform inputs. 
The limit of the mutual information of the A-channel when $K$ and $Q$ tend to infinity, with the ratio $\lambda = K/Q$ kept fixed, is studied in \cite{bassalygo2000calculation}.
Recent interest in the A-channel is due to the emergence of short-packet machine-type communication and IoT devices.
In particular, the idea of unsourced random access (URA) is introduced in \cite{polyanskiy2017perspective} as an alternative to the customary identify-then-transmit procedure. 
%
A common strategy in URA is to use concatenated coding techniques; when this strategy is adopted, in certain settings, the unsourced A-channel can serve as an abstraction for the channel between the outer encoders and the outer decoder. 
For example, in \cite{amalladinne2020coded}, a linear inner code is used to further encode outer-encoded messages, where the outer code allows for recovery over the unsourced A-channel. 
%
%
During inner decoding, each section is decoded independently, and a list of outer codeword symbols is produced for each section. 
The parity bits from the outer code designed for the unsourced A-channel are then used to link the outer codeword symbols corresponding to the same user across all the transmission sections.
The ensuing unsourced A-channel code is termed the tree code.
The original tree code has been enhanced in \cite{amalladinne2020coded,amalladinne2021unsourced}.
More recently, following the renewed interest in these applications of the A-channel, Lancho et al.~\cite{lancho2022finite} offered finite-blocklength achievability bounds for the unsourced A-channel.

\noindent
\textbf{Contributions:}
%
The major contributions of this article can be summarized as follows.
\begin{itemize}[nolistsep,leftmargin=*]
    \item  {\bf Sec.~\ref{sec:Channel Model} :} We introduce the UACE model, we define the information transmission problem for this model, and we identify relevant performance metrics. 
    %
    %
    
    \item {\bf Sec.~\ref{sec:Linked-loop code} :} We propose the linked-loop code
(LLC) as the first code specifically designed for the UACE. Encoding and decoding procedures are described. The code parameters are also discussed. 

\item {\bf Sec.~\ref{sec:Related Results} :} We review codes in the literature that are compatible with the UACE setting, although not specifically designed for this channel: (i) the tree code of \cite{amalladinne2020coded}, and the (ii) outer LDPC code (triadic design) of \cite{amalladinne2021unsourced}. 
%

\item {\bf Sec. \ref{sec:Simulation Results} : } Numerical simulations offer insight into the performance of the LLC in Sec.~\ref{sec:Linked-loop code}, and we provide a performance comparison with other systems in Sec.~\ref{sec:Related Results}.
\end{itemize}

{
\medskip
\noindent
{\bf Notation:} 
In the following, 
%
%
%
we adopt the short-hand notation 
$[N] \triangleq (0, \ldots, N-1)$. 
%
%
%
%
Calligraphic scripts are used to denote sets (e.g., $\Acal$), and $|\cdot|$ is used to denote the cardinality of a set.
$GF(2)$ indicates the Galois field of order 2.  
Given two row vectors $\vv$ and $\wv$, we denote their concatenation as $\vv.\wv$.
%
Random variables are indicated with uppercase letters (e.g., $N$). 
%
With a certain abuse of notation, we sometimes interchange the notation for (random) vectors and sets when this is clear from the context. 
%

\medskip
\noindent
The code for the numerical evaluations of this paper is provided online at \url{https://github.com/williamzheng0711/URAC23}.

\section{Channel Model}
\label{sec:Channel Model}

The UACE is a variation of the unsoured A-channel from \cite{lancho2022finite} that also encompasses symbol erasures.
%
%
In the (perfect) unsourced A-channel of \cite{lancho2022finite}, the channel output corresponding to the $l^{\rm th}$ channel use is defined as 
\ea{
\textstyle Y(l) = \bigcup_{k \in [K]} x_{k}(l),
\label{eq:channel perfect}
}
where $x_{k}(l) \in \Qcal$.
When transmission takes place over $L$ channel uses,  
we indicate the $L$ channel outputs as $\Yv=[ {Y}(0),\ldots,Y({L-1})]$ .
%
The $L$ channel inputs corresponding to user $k$ are labeled as $\xv_k=[ x_{k}(0),\ldots,x_{k}(L-1) ]$.  
%

The UACE is the variation of the model in \eqref{eq:channel perfect} defined in \eqref{eq:channel error}. 
%
Note that $|Y(l)|$ is a random variable which may be less than $K$ due to both collisions and erasures. 
Here, a collision refers to the event when two distinct users $i\neq j$ share a same symbol $x_{i}(l)=x_{j}(l) \in \Qcal$.
%
%
Note that the UACE in \eqref{eq:channel error} is parametrized by the number of users $K$, the cardinality of the input set $Q$, and the erasure probability $p_e$. 
%
%
%

\subsection{Binary Vector Representation}
\label{sec:Binary representation}

%
%
When describing coding schemes, it is often convenient to operate over a field; this is also the case also for the UACE.
For this reason, we introduce a specific notation for the case in which $Q=2^J$ for some $J \in \Nbb$. 
%
Let $x_{k}(l)$ be denoted by a $J$-bit binary vector $\vv_{k}(l)$. 
Then, the channel output may be expressed as
\ea{
\textstyle \yv(l) = \bigcup_{k \in [K]}  \delta_{E_{lk}} \lb \bold{v}_{k}(l) \rb.
}
After $L$ uses of the channel, what we obtain may be expressed as the tuple 
\ea{
\Yv =\lsb \yv(0),\yv(1), \ldots ,\yv({L-1})\rsb.
}
Moving forward, it shall be convenient to represent the concatenation of the $L$ input binary vectors, each comprised of $J$ bits, as a single row vector; that is
\ea{
\bold{v}_k = \bold{v}_k(0).\bold{v}_k(1) \ldots \bold{v}_k(L-1).
\label{eq:input binary size N}
}
In this case, we have that the binary representation of the input of user $k$ is  a binary vector $\vv_k$ of dimension $N=JL$.

%

%
%
%




\subsection{Coding for the UACE}

%
Let us consider the problem in which each user $k\in [K]$ wishes to transmit a payload of $B$ bits within $L$ channel uses. 
Assume payload $\Wv=\{\wv_k\}_{k \in [K]}$ is drawn uniformly at random from $[2]^B$. 
Let us refer to $\wv \in \Wv$ as an \emph{active} payload, whereas $[2]^B \setminus \Wv$ contains \emph{inactive} payloads.
Finally, if two users randomly select the same payload, let us refer to this event as a payload \emph{collision}.
\begin{definition}[$(L,B)$ Code]
\label{def:code}
A code of block-length $L$ and payload length $B$ for the UACE with parameters $(K,Q,p_e)$ is defined as a (possibly) random \emph{encoding} mapping to produce $L$ channel inputs at each user $k \in [K]$ to encode their message $\wv_k$, defined as
\ea{
\xv_k = f_{\rm enc} \lb  \wv_k \rb, 
\label{eq:encoding fun}
}
where $\wv_k \in [2]^B$ and $\xv_k \in \Qcal^L$,
together with a (possibly) random \emph{decoding} mapping which produces a list of transmitted payloads from the channel output, that is
\ea{
\Whv = 
f_{\rm dec}( \Yv),
\label{eq:decoding fun}
}
with  $|\Whv|=\Kh$.
\end{definition}
In other words, the encoding function maps the $B$-bit message $\wv_k$ into the $L$-tuple of channel inputs $\xv_k$. 
Upon observing the $L$ channel outputs, $\Yv$, the decoding function produces a list of payloads that are likely to be active. 
%
%
The list of possibly active payloads produced by the decoding function has length $\Kh$. 
%
%
%
%
Note that the rate of the code in Def.~\ref{def:code} is obtained as $R = B/L$. 
Also note that in a UACE each encoder $k \in [K]$ utilizes the same encoding function, so that the identity of the transmitter remains immaterial. 
%
%
\begin{definition}[Payload Probability of Error]
\label{def:pupe}
Given code with parameters $(N,B)$ for a UACE with parameters $(K,Q,p_e)$, define 
\begin{itemize}[nolistsep,leftmargin=*]
    \item the Payload Dropping Probability (PDP) as
    \ea{
P_{\rm PDP} = \f 1 K  \sum_{k \in [K]} \Pr \lsb \wv_k \not \in \Whv(\Yv) | \wv_k \in \Wv \rsb,
\label{eq:PDP}
}
    \item the Payload Hallucination Probability (PHP) as
\ea{
P_{\rm PHP} = \f 1 {\Kh}  \sum_{k' \in [\Kh]} \Pr \lsb  \wv_{k'} \not \in \Wv | \wv_{k'}  \in \Whv(\Yv) \rsb,
\label{eq:PHP}
}    
\end{itemize}
%
where the probability in \eqref{eq:PDP} and \eqref{eq:PHP} is taken over (i) 
the choice of the payload at each user which is selected uniformly at random over $[2]^B$,  (ii) the realization of the erasure sequence, and (iii) the possible randomness in the encoding and decoding functions. 
\end{definition}

More informally, the PDP indicates the probability that a payload is declared inactive when it was actually active, while the PHP indicates the probability that a payload is declared active when it was indeed inactive. 
The PDP and PHP are somewhat analogous to the type I and type II errors, respectively, in the hypothesis testing literature.

\subsection{Linear codes}
\label{sec:Linear codes}
%
%
%
%
Binary linear codes are a special case of the class of codes in Def. \ref{def:code} in which the encoding function of \eqref{eq:encoding fun} is a matrix multiplication over $GF(2)$.
More precisely, a binary linear code is defined as a set of $L$ 
\emph{generator matrices}
$\{\Gv_l\}_{l \in [L]}$  which specify how the parity bits for section $l \in [L]$ are generated from the payload $\wv$. 
Let $\{m(l)\}_{l\in [L]}$ with $\sum_{l \in [L]} m(l) = B$ indicate the number of information bits allocated to section $l \in [L]$, and recall that $\mv = [m(0),m(1), \ldots, m(L-1)]$
%
%
Next, we partition the payload $\wv$ into $L$ vectors,
\ea{
\wv = \wv(0) . \wv(1) \ldots  \wv(L-1),
\label{eq:payload partition}
}
with $\wv(l) \in [2]^{m(l)}$ for all $l\in [L]$.
Let the parity bits in section~$l$ be denoted by $\pv(l) \in \mathbb{F}_2^{p(l)}$, where length $p(l)=J-m(l)$. 
%
%
The parity bits are obtained from the payload as
\eas{
\pv(l) &= \wv \Gov_{l}  \\
&= \sum_{l' \in [L]} \wv(l') \Gv_{l'l},
\label{eq:parity equation 2}
}{\label{eq:parity equation}}
where $\Gov_{l} \in \mathbb{F}_2^{ B \times (J-m(l))}$.
When considering the partitioning of the payload as in \eqref{eq:payload partition}, one naturally obtains a partitioning of the $\Gov_{l}$ in sub-matrices  as in \eqref{eq:parity equation 2} where $\Gv_{l'l} \in \mathbb{F}_2^{m_{l'} \times (J - m(l))}$. 
%
%
Finally, the channel input is obtained as 
\ea{
\vv(l) = \wv(l).\pv(l).
}
The parameters found in this section are summarized in Tab.~\ref{tab:DNN parameters}.

\begin{figure}
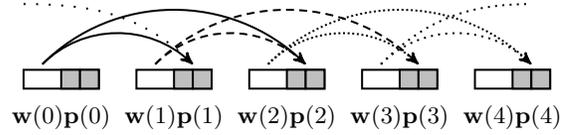

    \centering{\include{figures/codeword}}
    \vspace{-0.75cm}
    \caption{This notional diagram depicts the formation of a linked-loop codeword.}
    \label{fig:codeword}
\end{figure}

\begin{remark}
Throughout the section, we have omitted the subscript $k$ found in \eqref{eq:input binary size N}.
This simplifies notation and it stresses the fact that the encoding is the same at all users $k \in [K]$.
\end{remark}

\begin{table}
	\footnotesize
	\centering
	\vspace{0.04in}\caption{Relevant Symbols and Parameters}
	\label{tab:DNN parameters}
	\begin{tabular}{|c|c|}
		\hline
        Notation & Parameter Description \\ \hline
        $B$ & Message length in bits \\ \hline
        $L$ & Number of coded sub-blocks per round \\\hline
        $J$ & Length of coded sub-blocks  \\ \hline
        $N$ & Length of coded $B$-bit message, $N=LJ$ \\ \hline
        $M$ & Number of previous sub-blocks used to generate parity bits\\ \hline
        $m(l)$ & Number of information bits in $l$-th sub-block \\ \hline
        $p(l)$ & Number of parity bits in $l$-th sub-block, $p(l) = J-m(l)$ \\ \hline
	\end{tabular}
\end{table}

\section{Linked-loop code (LLC)}
\label{sec:Linked-loop code}

We now discuss the task of correctly identifying active payloads before introducing the LLC approach. 
For simplicity, let us consider the regime in which each user experiences no erasures.
%
Also, we will assume no payload collision occurs, as the probability of this event is orders of magnitude smaller that the target PDP and PHP.

The decoder begins by selecting a single entry of $\yv(0)$, and sets this entry as the root of a tree. 
Then, the decoder checks all entries in $\yv(1)$ and identifies those entries that are parity consistent with the root entry. 
For each such entry, the decoder creates a path from section $0$ to section~$1$. 
%
%
Proceeding, the decoder checks all entries in $\yv(2)$ and identifies those entries that are parity consistent with the previously identified partial paths through $\yv(0)$ and $\yv(1)$. 
Retaining only parity-consistent paths, the decoder proceeds until all sections have been considered. 
In the tail-biting scenario, the first sections may need to be re-considered to ensure that they are parity-consistent with the last sections. 
The hope is that, by the end, a single parity-consistent path remaining, which corresponds to a transmitted codeword. 
As in \cite{amalladinne2020coded}, we refer to this operation as \emph{stitching} and illustrate the algorithm in Fig.~\ref{fig:stitching}.

Next, let us discuss a decoding approach when only one erasure is encountered. 
When an erasure occurs for a certain user, the decoder is not able to insert the correct element in one of the lists, let's say $\yv_l$. 
Consequently, the number of surviving paths becomes zero for some section $l'$ with $l \leq l' \leq L$. 
This means that, after all initial paths have been linked, some of the elements of the root do not have a complete path of length $L$. For these remaining codewords, a \emph{phase 2} of decoding is needed in which all non-terminating paths are considered.
For these paths, we introduce a special symbol, $\na$, that enables each path to have a one-section loss tolerance. 
After all other sections not in outage have been stitched to some path, we calculate the lost section's information bits by reverse-engineering \eqref{eq:parity equation}.

\begin{figure}
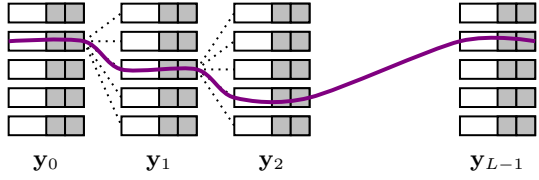

    \centering{\include{figures/stitching}}
    \vspace{-0.75cm}
    \caption{The diagram illustrates the stitching of elements across sections.
    When successful, this process returns candidate codewords.}
    \vspace{-0.25cm}
    \label{fig:stitching}
\end{figure}

\subsection{Encoding}
The LLC is a binary linear code for the UACE defined by: (i) a memory depth $M$,  and (ii)  a set of $M$ parity matrices $\{\Gv_r 
\}_{r = 1}^{M}$.
The LLC has a fixed number of information and parity bits in each section, i.e.,
$m(l)=m$ for some $m>0$.
Additionally, the $J-m$ parity bits depend on the information bits of the $M$ previous sections, in a circular manner, that is 
\eas{
\pv (l) & = \textstyle  \sum_{l' \in [L]} \wv(l')\Gv_{l'l} \\
& = \textstyle \sum_{r =1}^{M} \wv( l-r \, {\rm mod} \, L) \Gv_{r},
\label{eq:parity equation 3}
}
where in \eqref{eq:parity equation 3} we have set $\Gv_{l'l}=\Gv_{r}$ for $r = l-l' \, {\rm mod} \, L$ and $1\leq r\leq M$.



For example, if $L=16$ and $M=2$, the first three lists of parity bits are obtained as:
\eas{
\pv(0) & =  \wv(14)\Gv_{2} + \wv(15)\Gv_{1} \\
\pv(1) & =  \wv(15)\Gv_{2} + \wv(0)\Gv_{1} \\
\pv(2) & =  \wv(0)\Gv_{2} + \wv(1)\Gv_{1}.
}
%


\subsection{Decoding}

For the sake of compactness, we present the decoding algorithm through pseudo-code in Alg.~\ref{alg:cap}. 
Therein, $\textbf{checkParity}(\uv)$ simply checks whether the parity bits of all sections of the input $\uv$ are consistent with \eqref{eq:parity equation 3}; thus, it returns $\textbf{False}$ if and only if \eqref{eq:parity equation 3} is violated with all involved sections that are present as explicit bits in $\uv$. 
Hence when a path is of length $\leq M$, it will always be parity-consistent because no section's parity can be checked at all. 
The subroutine $\textbf{checkParityWithRcv}(\uv)$ serves a similar purpose, except for when $\uv$ contains an $\na$, it tries to uniquely determine the lost information bits according to \eqref{eq:parity equation 3} by reverse-engineering. 
This can be done when we choose matrices in $\{\Gv_r\}_{r=1}^M$ that admit a right-inverse.

\begin{algorithm}[H]
\caption{Pseudo code of the LLC decoding algorithm.}
\label{alg:cap}
\begin{algorithmic}[1]
\Require UACE output $\Yv=[\yv(0), \ldots , \yv(L-1)]$ 
\Ensure 
$\forall \ l \in [L]$, $|\yv(l)| \leq K$; 

\Comment Decoding phase $1$

\State $\Lsf^1 \gets \emptyset$; $\widehat{\Wv}^1 \gets \emptyset$

\ForEach{$l \in [L]$}
    \ForEach{$\uv \in (\Lsf^1 \text{ if } \Lsf^1\neq \emptyset \text{ else } \{\emptyset\})$}
        \State $\Lsf^1 \gets \Lsf^1 - \{\uv\}$
        \ForEach{$x(l) \in \yv(l)$}
            \If{\textbf{checkParity}$(\bold{u}. x(l))$}
                \State $\Lsf^1 \gets \Lsf^1 \cup \{\uv.x(l)\}$
            \EndIf
        \EndFor
    \EndFor
\EndFor

\State $\widehat{\Wv}^1 \gets \textbf{extractInfoBits}(\Lsf^1)$

\Comment{Decoding phase $2$}

\State $\Lsf^2 \gets \emptyset$; $\widehat{\Wv}^2 \gets \emptyset$

\ForEach{$x(0)^\times \in \yv(0)$ leads to no element in $\Lsf^1$}
    \State $\Lsf^2(x(0)^\times) \gets \{x(0)^\times \}$
    \ForEach{$l \in \{1,2,\ldots,L-1\}$}
        \ForEach{$\uv \in \Lsf^2(x(0)^\times)$}
            \State $\Lsf^2(x(0)^\times) \gets \Lsf^2(x(0)^\times) - \{\uv\}$
            \If{$\uv \text{ contains no } \na$ }
                \State $\Lsf^2(x(0)^\times) \gets \Lsf^2(x(0)^\times) \cup \{\uv . \na \}$

            \EndIf
            \ForEach{$x(l) \in \yv(l)$}
                \If{$\textbf{checkParityWithRcv}(\uv . x(l))$}
                    \State $\Lsf^2(x(0)^\times) \gets \Lsf^2(x(0)^\times) \cup \{\uv . x(l)\}$
                \EndIf
            \EndFor
        \EndFor
    \EndFor

    \State $\Lsf^2 \gets \Lsf^2 \cup \Lsf^2(x(0)^\times)$
\EndFor


\State $\widehat{\Wv}^2 \gets \textbf{extractInfoBits}(\Lsf^2)$

\State $\widehat{\Wv} \gets \widehat{\Wv}^1 \cup \widehat{\Wv}^2$

\State \Return $\widehat{\Wv}$


\end{algorithmic}
\end{algorithm}

\section{ Related Results}
\label{sec:Related Results}
In our numerical simulations, we shall compare the performance of the LLC with the tree code from \cite{amalladinne2020coded} and the LDPC code from \cite{amalladinne2021unsourced}.
We stress that both of these codes were not originally designed for the UACE model and, as such this comparison offers only limited insight.
These schemes are selected as benchmarks because they are readily applicable to the problem at hand.

%

\noindent
{\bf Tree code (TC):}
The tree code of \cite{amalladinne2020coded} is a binary linear code in which the length of parity bits, $p(l)$, increases by each section. 
In decoding, it is assumed that $Y$ in \eqref{eq:channel error} does not contain any erasures but possibly contains additional, untransmitted symbols.
%
For this reason, the decoding algorithm is substantially equivalent to phase~$1$ in the LLC decoding, with the difference that the number of parity bits increases with the section number, thus allowing for a more efficient pruning of surviving paths.

\noindent
{\bf LDPC:}
Another candidate coding scheme for the UACE is the triadic LDPC code with belief propagation (BP) decoding of \cite{amalladinne2021unsourced}.
In this approach, sections are viewed as symbols over a Galois field and they are encoded using a non-binary LDPC code; therein, sections contain either all information bits or all parity bits. 
Parity sections are obtained from information sections as in a classic LDPC code. 
The FFT method for decoding non-binary LDPC codes is leveraged to efficiently implement message passing between parity sections and information sections.

%
%

\section{Simulation Results}
\label{sec:Simulation Results}
In this section, we study the performance of the LLC on the UACE with hyperparameters tuned to practical applications. 
We provide a comparison with the tree code of \cite{amalladinne2020coded} and the LDPC code of \cite{amalladinne2021unsourced}. 
We consider a system with $K$ active users, each having $B=128$ bits of information to transmit. 
The corresponding codewords are divided into $L=16$ sub-blocks, each of length $J=16$. 
For the tree code, the specific value for $m(l)$ leads to different discriminating power and was chosen carefully  after extensive numerical experimentation. 
Comparison of the three codes using fixed $K=100$ and varying $p_e \in [0,0.15]$ is shown in Fig.~\ref{fig:llc_vs_tc_vs_ldpc}. Compared to TC and LDPC code, LLC offers a noticeable improvements in terms of greatly decaying the PDP due to phase 2 of decoding, while keeping PHP in a low level. Fig.~\ref{fig:only_llc} shows how LLC itself performs under different $(K,p_e)$ with $K\in [50,150]$.

\section{Conclusion}
This paper presents a novel coding scheme designed for UACE called linked-loop code (LLC). The LLC is managed to identify all payloads suffer no erasure, also to recover payloads suffer one erasure from the UACE. The latter greatly decreases PDP, while keeping PHP to be empirically low comparing with some other methods in the recent literature.

\begin{figure}
  \centering
  \input{figures/llc_vs_tc_vs_ldpc_performance}
  \caption{Empirical $P_{\text{PHP}}$ vs $P_{\text{PDP}}$ for the LLC (with $M=2$), tree code, and LDPC code. Here, $K = 100$, and the $p_e$ associated with each point is labelled on the graph.
  }
 \label{fig:llc_vs_tc_vs_ldpc}
\end{figure}
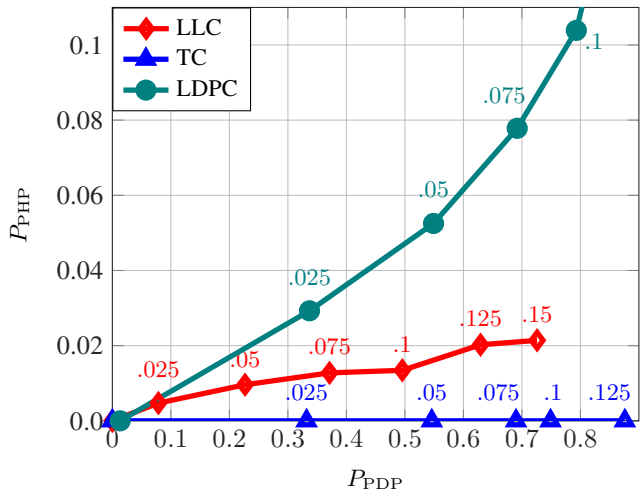

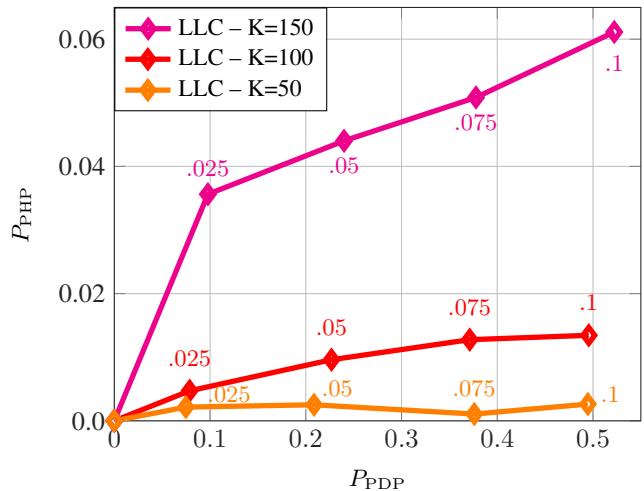
\begin{figure}
  \centering
  \input{figures/only_llc_changing_kpe}
  \caption{Empirical $P_{\text{PHP}}$ vs $P_{\text{PDP}}$ of the LLC (with $M=2$) for various values of $K$ and $p_e$. The label associated with each data point indicates the $p_e$. }
 \label{fig:only_llc}
\end{figure}


\bibliographystyle{IEEEbib}
\bibliography{IEEEabrv,A_channel}
\end{document}

%% file: preamble.tex
\usepackage[utf8]{inputenc}

\usepackage{amsmath}
\usepackage{amsfonts}
\usepackage{amssymb}
\usepackage{textcomp}
\usepackage{color}
\usepackage{xcolor}
\usepackage{amsthm}
\usepackage{bm}
\usepackage{mathrsfs}

\usepackage{algorithmicx}
\usepackage{algorithm}
\usepackage[noend]{algpseudocode}
\usepackage{soul}
\usepackage{graphicx}
\usepackage{subcaption}
\usepackage{epsf}
\usepackage{xpatch}
\usepackage{setspace} 
\usepackage{enumitem,kantlipsum}
\usepackage{multirow,booktabs}
\usepackage{latexsym}
\usepackage{hyperref}
\usepackage{cleveref}
\usepackage[export]{adjustbox}
\usepackage{dsfont}
\usepackage{mathcomSTEv4}
\usepackage{tikz}
\usepackage{circuitikz}
\usetikzlibrary{arrows,shapes}
\usepackage{mathtools}

\theoremstyle{definition}
\newtheorem{definition}{Definition}[section]

\newtheorem*{remark}{Remark}

\algnewcommand\algorithmicforeach{\textbf{for each}}
\algdef{S}[FOR]{ForEach}[1]{\algorithmicforeach\ #1\ \algorithmicdo}
\usetikzlibrary{decorations.pathreplacing,calc}

%% file: figures/codeword.tex
\begin{tikzpicture}
[font=\small, draw=black,  >=stealth', line width=0.75pt,
sub0/.style={rectangle, draw, inner sep=0pt, minimum width=10mm, minimum height=2.5mm},
parity/.style={rectangle, draw, fill=lightgray, inner sep=0pt, minimum size=2.5mm}]

\foreach \c in {0.00} {
  \node[sub0] (subcs1\c) at (-1.00,\c) {};
  \node[parity] (parity1a\c) at (-0.875,\c) {};
  \node[parity] (parity1b\c) at (-0.625,\c) {};
  \node[sub0] (subcs2\c) at (0.50,\c) {};
  \node[parity] (parity2a\c) at (0.625,\c) {};
  \node[parity] (parity2b\c) at (0.875,\c) {};
  \node[sub0] (subcs3\c) at (2.00,\c) {};
  \node[parity] (parity3a\c) at (2.125,\c) {};
  \node[parity] (parity3b\c) at (2.375,\c) {};
  \node[sub0] (subcs4\c) at (3.50,\c) {};
  \node[parity] (parity4a\c) at (3.625,\c) {};
  \node[parity] (parity4b\c) at (3.875,\c) {};
  \node[sub0] (subcs5\c) at (5.00,\c) {};
  \node[parity] (parity5a\c) at (5.125,\c) {};
  \node[parity] (parity5b\c) at (5.375,\c) {};
}

\draw [->] (-1.25,0.2) to [out=45,in=135] (0.75,0.2);
\draw [->] (-1.25,0.2) to [out=45,in=135] (2.25,0.2);
\draw [->,densely dashed] (0.25,0.2) to [out=45,in=135] (2.25,0.2);
\draw [->,densely dashed] (0.25,0.2) to [out=45,in=135] (3.75,0.2);
\draw [->, densely dotted] (1.75,0.2) to [out=45,in=135] (3.75,0.2);
\draw [->, densely dotted] (1.75,0.2) to [out=45,in=135] (5.25,0.2);
\draw [->, dotted] (3.25,0.2) to [out=45,in=135] (5.25,0.2);
\draw [-, dotted] (3.25,0.2) to [out=45,in=180] (5.25,1);
\draw [->, loosely dotted] (-1.50,1) to [out=0,in=135] (0.75,0.2);

\node (info0) at (-1,-0.5) {$\wv(0)\pv(0)$};
\node (info1) at (0.5,-0.5) {$\wv(1)\pv(1)$};
\node (info2) at (2,-0.5) {$\wv(2)\pv(2)$};
\node (info3) at (3.5,-0.5) {$\wv(3)\pv(3)$};
\node (infoz) at (5.00,-0.5) {$\wv(4) \pv(4)$};

\end{tikzpicture}

%% file: figures/stitching.tex
\begin{tikzpicture}
[font=\small, draw=black, line width=0.75pt,
sub0/.style={rectangle, draw, inner sep=0pt, minimum width=10mm, minimum height=2.5mm},
parity/.style={rectangle, draw, fill=lightgray, inner sep=0pt, minimum size=2.5mm}]

\foreach \c in {3.00, 2.625, 2.25, 1.875, 1.5} {
  \node[sub0] (subcs1\c) at (-1.00,\c) {};
  \node[parity] (parity1a\c) at (-0.875,\c) {};
  \node[parity] (parity1b\c) at (-0.625,\c) {};
  \node[sub0] (subcs2\c) at (0.50,\c) {};
  \node[parity] (parity2a\c) at (0.625,\c) {};
  \node[parity] (parity2b\c) at (0.875,\c) {};
  \node[sub0] (subcs3\c) at (2.00,\c) {};
  \node[parity] (parity3a\c) at (2.125,\c) {};
  \node[parity] (parity3b\c) at (2.375,\c) {};
  \node[sub0] (subcsz\c) at (5.00,\c) {};
  \node[parity] (parityza\c) at (5.125,\c) {};
  \node[parity] (parityzb\c) at (5.375,\c) {};
}

\draw[dotted] (-0.50,2.625) -- (0.00,3.00) {};
\draw[dotted] (-0.50,2.625) -- (0.00,2.625) {};
\draw[dotted] (-0.50,2.625) -- (0.00,2.25) {};
\draw[dotted] (-0.50,2.625) -- (0.00,1.875) {};
\draw[dotted] (-0.50,2.625) -- (0.00,1.50) {};

\draw[dotted] (1.00,2.25) -- (1.50,3.00) {};
\draw[dotted] (1.00,2.25) -- (1.50,2.625) {};
\draw[dotted] (1.00,2.25) -- (1.50,2.25) {};
\draw[dotted] (1.00,2.25) -- (1.50,1.875) {};
\draw[dotted] (1.00,2.25) -- (1.50,1.50) {};

\node (list1) at (-1.00,1) {$\yv_0$};
\node (list2) at (0.50,1) {$\yv_1$};
\node (list3) at (2.00,1) {$\yv_2$};
\node (list4) at (5.00,1) {$\yv_{L-1}$};

\draw [line width=1.5pt,color=violet] plot[smooth, tension=.5] coordinates {
(-1.50,2.625) (-0.50,2.625)
(0.00,2.25) (1.00,2.25)
(1.50,1.875) (2.50,1.875)
(4.50, 2.625) (5.50, 2.625)};
\end{tikzpicture}

%% file: figures/llc_vs_tc_vs_ldpc_performance.tex
\pgfplotsset{scaled y ticks=false}
\begin{tikzpicture}

\begin{axis}[
    font=\small,
    width=7cm,
    height=5.5cm,
    scale only axis,
    every outer x axis line/.append style={white!15!black},
    every x tick label/.append style={font=\color{white!15!black}},
    xmin=0,
    xmax=0.9,
    xtick = {0.0, 0.1, ..., 0.9},
    xlabel={$P_{\mathrm{PDP}}$},
    xmajorgrids,
    every outer y axis line/.append style={white!15!black},
    every y tick label/.append style={font=\color{white!15!black}},
    ymin=0,
    ymax=0.11,
    ytick = {0.0, 0.02, ..., 1.0},
    yticklabels={0.0, 0.02, 0.04, 0.06, 0.08, 0.1, 0.12, 0.14},
    ylabel={$P_{\mathrm{PHP}}$},
    ymajorgrids,
    legend style={at={(0,1)},anchor=north west, draw=black,fill=white,legend cell align=left}
]

\addplot [
    color=red,
    solid,
    line width=2.0pt,
    mark size=3.0pt,
    mark=diamond,
    mark options={solid}
]
table[row sep=crcr]{
    0.0 0.0 \\
    0.07875 0.00472 \\
    0.226875 0.009607 \\
    0.37125 0.0127576 \\
    0.495625 0.013447 \\
    0.629375 0.020236 \\
    0.726 0.02142 \\
};
\addlegendentry{LLC};

\node[label={90:{\color{red}$.025$}}] at (axis cs:0.07875, 0.00672) {};
\node[label={90:{\color{red}$.05$}}] at (axis cs:0.226875, 0.009407) {};
\node[label={90:{\color{red}$.075$}}] at (axis cs:0.37125, 0.0127576) {};
\node[label={90:{\color{red}$.1$}}] at (axis cs:0.495625, 0.013447) {};
\node[label={90:{\color{red}$.125$}}] at (axis cs:0.629375, 0.020236) {};
\node[label={90:{\color{red}$.15$}}] at (axis cs:0.726, 0.02142) {};

\addplot [
    color=blue,
    solid,
    line width=2.0pt,
    mark size=3.0pt,
    mark=triangle,
    mark options={solid}
]
table[row sep=crcr]{
    0.0 0.0 \\
    0.332 0 \\
    0.546 0 \\
    0.69 0 \\
    0.749 0 \\
    0.876 0 \\
};
\addlegendentry{TC};

\node[label={90:{\color{blue}$.025$}}] at (axis cs:0.332, 0.0005) {};
\node[label={90:{\color{blue}$.05$}}] at (axis cs:0.546, 0.0005) {};
\node[label={90:{\color{blue}$.075$}}] at (axis cs:0.66, 0.0005) {};
\node[label={90:{\color{blue}$.1$}}] at (axis cs:0.753, 0.0005) {};
\node[label={90:{\color{blue}$.125$}}] at (axis cs:0.85, 0.0005) {};

\addplot [
    color=teal,
    solid,
    line width=2.0pt,
    mark size=3.0pt,
    mark=otimes*,
    mark options={solid}
]
table[row sep=crcr]{
    0.01333 0.0 \\
    0.337 0.02928 \\
    0.549 0.05252 \\
    0.692 0.077844 \\
    0.793 0.103896 \\
    0.876 0.15646  \\
};
\addlegendentry{LDPC};

\node[label={90:{\color{teal}$.025$}}] at (axis cs:0.337, 0.03128) {};
\node[label={90:{\color{teal}$.05$}}] at (axis cs:0.549, 0.05452) {};
\node[label={90:{\color{teal}$.075$}}] at (axis cs:0.67, 0.079644) {};
\node[label={90:{\color{teal}$.1$}}] at (axis cs:0.823, 0.093896) {};

\end{axis}

\end{tikzpicture}

%% file: figures/only_llc_changing_kpe.tex
\pgfplotsset{scaled y ticks=false}
\begin{tikzpicture}

\begin{axis}[
    font=\small,
    width=7cm,
    height=5.5cm,
    scale only axis,
    every outer x axis line/.append style={white!15!black},
    every x tick label/.append style={font=\color{white!15!black}},
    xmin=0,
    xmax=0.55,
    xtick = {0.0, 0.1, ..., 0.9},
    xlabel={$P_{\mathrm{PDP}}$},
    xmajorgrids,
    every outer y axis line/.append style={white!15!black},
    every y tick label/.append style={font=\color{white!15!black}},
    ymin=0,
    ymax=0.065,
    ytick = {0.0, 0.02, ..., 1.0},
    yticklabels={0.0, 0.02, 0.04, 0.06, 0.08, 0.1},
    ylabel={$P_{\mathrm{PHP}}$},
    ymajorgrids,
    legend style={at={(0,1)},anchor=north west, draw=black,fill=white,legend cell align=left}
]

\addplot [
    color=magenta,
    solid,
    line width=2.0pt,
    mark size=3.0pt,
    mark=diamond,
    mark options={solid}
]
table[row sep=crcr]{
    0.0 0.0 \\
    0.09777777 0.03562945 \\
    0.24 0.04402515723 \\
    0.37777777 0.05084745763 \\
    0.52222222 0.06113537118 \\
};
\addlegendentry{LLC -- K=150};

\node[label={90:{\color{magenta}$.025$}}] at (axis cs:0.09778, 0.03562945) {};
\node[label={90:{\color{magenta}$.05$}}] at (axis cs:0.24, 0.03602515723) {};
\node[label={90:{\color{magenta}$.075$}}] at (axis cs:0.37778, 0.04284745763) {};
\node[label={90:{\color{magenta}$.1$}}] at (axis cs:0.5222, 0.05213537118) {};

\addplot [
    color=red,
    solid,
    line width=2.0pt,
    mark size=3.0pt,
    mark=diamond,
    mark options={solid}
]
table[row sep=crcr]{
    0.0 0.0 \\
    0.07875 0.00472 \\
    0.226875 0.009607 \\
    0.37125 0.0127576 \\
    0.495625 0.013447 \\
};
\addlegendentry{LLC -- K=100};

\node[label={90:{\color{red}$.025$}}] at (axis cs:0.07875, 0.00572) {};
\node[label={90:{\color{red}$.05$}}] at (axis cs:0.226875, 0.010607) {};
\node[label={90:{\color{red}$.075$}}] at (axis cs:0.37125, 0.0137576) {};
\node[label={90:{\color{red}$.1$}}] at (axis cs:0.495625, 0.014447) {};

\addplot [
    color=orange,
    solid,
    line width=2.0pt,
    mark size=3.0pt,
    mark=diamond,
    mark options={solid}
]
table[row sep=crcr]{
    0.0 0.0 \\
    0.07466 0.002156 \\
    0.20866 0.002521 \\
    0.376 0.001067 \\
    0.49466 0.00263 \\
};
\addlegendentry{LLC -- K=50};

\node[label={90:{\color{orange}$.025$}}] at (axis cs:0.12, 0) {};
\node[label={90:{\color{orange}$.05$}}] at (axis cs:0.233, 0.001) {};
\node[label={90:{\color{orange}$.075$}}] at (axis cs:0.376, 0.001) {};
\node[label={90:{\color{orange}$.1$}}] at (axis cs:0.518, 0) {};

\end{axis}

\end{tikzpicture}